# Feasibility Study of a Hybrid Solid–Liquid Vibration Energy Harvester – Numerical Simulation & Analysis

**Nadish Anand & Warren Jasper**

Mechanical & Aerospace Engineering, College of Engineering & Textiles Engineering
Chemistry & Science (TECS), College of Textiles
North Carolina State University, Raleigh, NC 27695

**Abstract**: In this paper, we have introduced and studied the feasibility of a hybrid solid-liquid vibration energy harvester. The energy harvester consists of a ferrofluid partially filled in a tank and a piezoelectric beam fixed at one of the tank walls. The tank is assumed to be placed in a non-uniform magnetic field created by placing two powerful magnets symmetrically external to the tank walls. This magnetic field was then implemented by a Magnetic field function modeling the magnetic field produced by the two magnets. We used piezoelectric beam configurations and oscillation loads to study and characterize this 2-D multiscale, multiphysics, and multiphase fluid-structure interaction. The tank is subjected to an external oscillatory motion, which sloshes the ferrofluid and oscillates the beam utilizing two modes: one due to the beam's inertia and the second due to the impact of the ferrofluid on the beam. The parameters varied in piezo materials, ferrofluid fill height, piezoelectric beam length, and oscillation frequency. The natural frequency modes for the piezoelectric beam are very important since the beam harvests the highest power in those modes. We have observed that when the fluid motion vibrates the beam, in some instances, the voltage output from piezoelectric material peaks to a high value, which is indicative of material nearing resonance frequencies at the fluid loading realized in those instances. However, in other cases, the voltage output varies with the loading, and hence, the response of the piezoelectric beam is broadband, very similar to the sloshing, which is inherently broadband.

Corresponding Author: Nadish Anand, Email: nanand@alumni.ncsu.edu

# 1. Introduction

Ferrofluids are colloidal suspensions of Ferromagnetic particles, often coated by an organic compound, suspended in a carrier fluid[1]. These carrier fluids could be anything from Deionized Water to mineral oils, etc., exhibiting varying thermophysical properties[1] Ferrofluids exhibit the unique property of super-paramagnetism, which allows them to exhibit bulk magnetization when an external magnetic field is applied. The bulk magnetization results from the alignment of the magnetic dipoles in the ferromagnetic particles inside the ferrofluid in the direction of the externally applied magnetic fields. Therefore, due to this external tunability of the ferrofluids, they are primarily used in flow manipulation applications such as rheology, heat transfer enhancement, mixing enhancement, energy harvesting, sensing, refrigeration, etc.[2], [3], [4]

Vibration energy harvesting (VEH) systems convert ambient vibrations into electric power. The primary purpose of the VEH systems is to provide seamless power and eventually replace the need for batteries in electronic devices. Most VEH systems consist of an external excitation coupled to a transduction mechanism, which supplies power to an external circuit. In most cases, the transduction mechanism includes a solid-state transducer, which transforms the deformations into electric charge, for example, piezoelectric devices, electroactive polymers, etc. [5], [6], [7]. These solid-state VEH systems, however, have a significant disadvantage: the energy harvesting potential is very "narrow band" (i.e., the highest output from these devices is achieved at or around resonant frequency modes). The incongruency between ambient vibrations, which usually are very wide-band, and the resonant modes further exacerbate the deficiencies in energy harvesting realized versus the potential. Recently, a novel VEH system was introduced with a ferrofluid as a liquid-state transduction mechanism[8], which was discussed in earlier sections. This system has a significant advantage over solid-state VEH systems, as the resonant modes for 'sloshing' are tightly distributed. Thus, it is more suited for harvesting energy from wide-band ambient vibrations[9], [10]. However, the power output from such a system depends heavily upon magnetic parameters[3], [11], [12] and thus can have limitations on their power output. Ferrofluid Sloshing Vibration Energy harvesting systems have gained attention recently, and several works have been published exploring different sloshing configurations subjected to varying magnetic field configurations[12], [13], [14], [15]. The idea is to produce a varying magnetic flux density inside a partially filled tank of ferrofluid, subjected to oscillatory impulse inside a spatially non-uniform magnetic field. This time-varying magnetic flux density can then be induced as current in a pickup coil wound on the tank.

This work postulates a combined sloshing and piezoelectric transduction vibration energy harvester. This system is then studied numerically to assess its feasibility. The principle behind such postulation is to utilize the kinetic energy of the ferrofluid not only as a liquid state transduction mechanism but also as a liquid-solid coupled system with a piezoelectric as an additional transduction mechanism. The hybrid energy harvester setup is discussed in the next section, and the third section presents details about governing equations and physical properties of the materials. In the fourth section, the numerical modeling details are presented, and the fifth section discusses the results of the numerical analysis of the system.

## 2. Problem setup

**Figure 1** shows the schematic representation of the hybrid energy harvester design used in this study. The design consists of a tank of length $L$ and height $H$ partially filled with a ferrofluid to a height $h$. On the top wall of the tank, a piezoelectric bimorph is affixed. The bimorph has piezoelectric beams of thickness $t_p$ and height $h_p$ attached to the base end of a steel beam of thickness $t_b$ and height $h_b$. The midplane of the bimorph coincides with the midplane of the tank.

The piezoelectric bimorph has two piezoelectric patches of smaller length than the central beam attached at the base. The central beam is of structural steel and is used as a terminal for voltage produced in the piezo material.

Similarly, for the ferrofluid, the electromotive force generated is assumed to be picked up by an external pick-up coil with 2000 turns. For simplification, the applied magnetic field over the harvester area is assumed to be unidirectional in the x-direction [16]. This system is excited through periodic lateral external excitation.

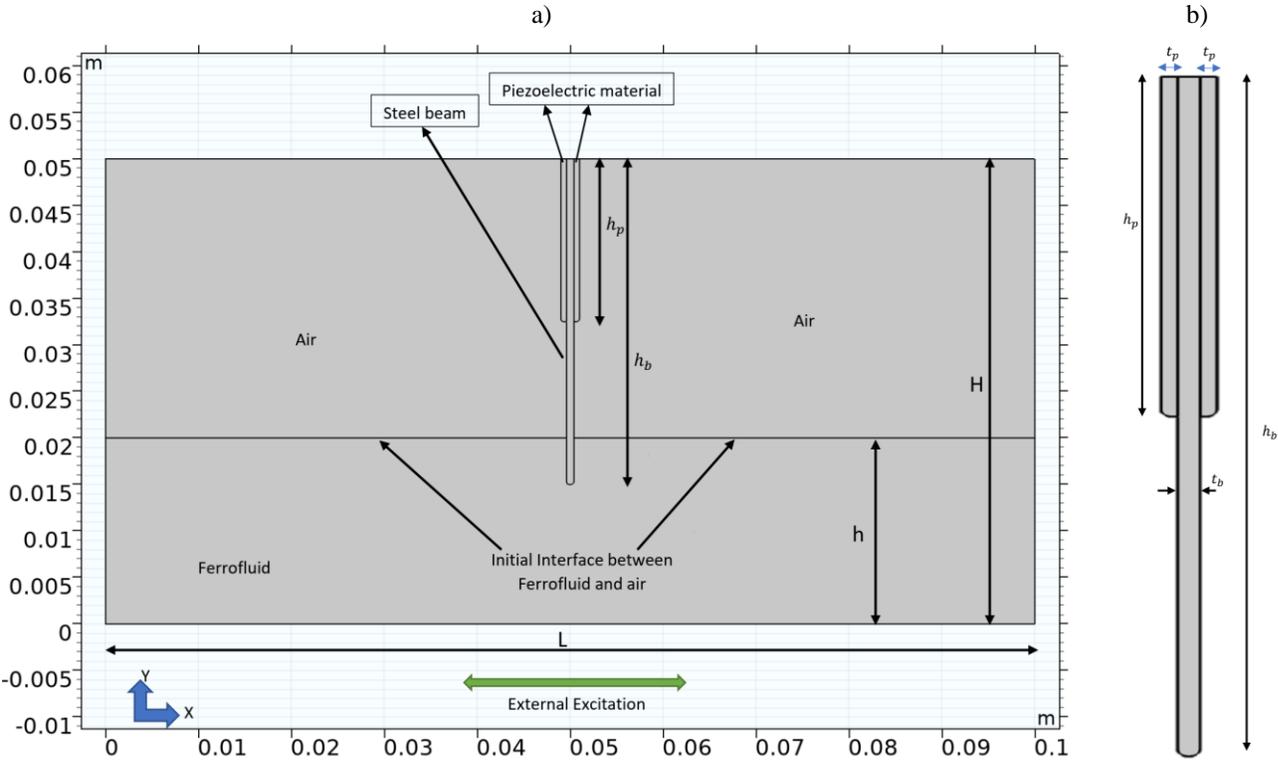

**Figure 1** Configuration of the hybrid VEH system a) the tank and piezo bimorph and b) close-up of the piezo bimorph

The applied magnetic field applied to the harvester is shown in **Figure 2**.

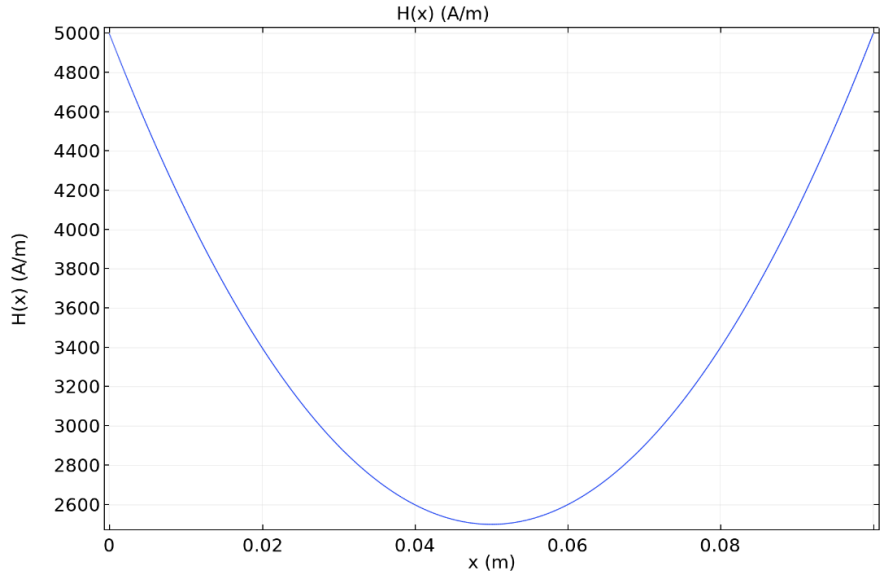

**Figure 2** Magnetic Field function

## 3. Governing Equations

The proposed system in the previous section would be governed by different physical phenomena inter-twined with each other. This makes the above system a multiphysics system. The following phenomena are observed:

1) Fluid mechanics (Navier-Stokes)
2) Free surface flow (Surface tension – Level set)
3) Wall wetting (Contact angle [17])
4) Fluid-structure interaction
5) Ferrohydrodynamics (Maxwell's equations)
6) Electrodynamics in piezoelectric (Maxwell's equations), and
7) Finally, a moving mesh interface is used to track the deformation of the bimorph.

Here, the fluid mechanics of the ferrofluid is influenced by the Kelvin body force exerted by the magnetic field (or magnets[12]) on the ferrofluid. Secondly, the fluid flow interacts with the piezoelectric bimorph and displaces the beam, influencing the flow (fully coupled). Thirdly, the free surface changes form; hence, the interface between the air and the fluid keeps moving and needs to be tracked to obtain the correct distribution of force fields inside the harvester. Fourthly, the stresses induced in the bimorph transform into strains in the piezoelectric material, which gives rise to charge displacement and, hence, a voltage/current. Finally, the ferrofluid wets the wall and inner surfaces; hence, the surface tension force changes. Wetting involves complicated physics and needs to be appropriately accounted for to define the correct surface profile between ferrofluid and air[12], [18], [19]. The governing equations pertaining to the above physical phenomena and their couplings are defined in the equations below.

a) Fluid Mechanics

Continuity:

$$\rho \nabla . \vec{V} = 0 \tag{1}$$

Momentum:

$$\rho \frac{\partial \vec{V}}{\partial t} + \rho (\vec{V}.\nabla)\vec{V} = -\nabla p + \mu \nabla^2 \vec{V} + \overrightarrow{F_m} + \rho \vec{g} + \overrightarrow{F_{st}} + \overrightarrow{F_{FSI}} \tag{2}$$

where, $\vec{V}$ is the velocity vector, $\overrightarrow{F_m}$ is the magnetic force, $\vec{g}$ is the external acceleration on the system, including the external excitation, $\overrightarrow{F_{st}}$ is the force of the surface tension and $\overrightarrow{F_{FSI}}$ is the force applied by the piezoelectric bimorph beam on the fluid elements.

Here the external excitation is given as:

$$x = X_0 Sin(\omega t) \tag{4}$$

where $X_0$ is the amplitude of external excitation and $\omega$ the angular frequency, which is related to the excitation frequency $f$ as, $2\pi f$. In this study, this $f$ is varied from 0.8 to 1.2 Hz to simulate low-frequency high-amplitude excitation.

b) Level-set equations for interface tracking:

$$\frac{\partial \phi}{\partial t} + \vec{V}.\nabla \phi = \gamma \nabla.\left(\epsilon_{ls}\nabla \phi - \phi(1-\phi)\frac{\nabla \phi}{|\nabla \phi|}\right) \quad (5)$$

where, $\phi$ is the level set variable, $\gamma$ is the re-initialization parameter and $\epsilon_{ls}$ is the parameter that controls the thickness of the fluid-air interface.

c) The equations for flow coupling of magnetic force and surface tension force are as follows

$$\overrightarrow{F_{st}} = \sigma_m \delta \kappa \boldsymbol{n_i} + \delta \nabla_s \sigma_m \quad (6)$$

where, $\sigma_m$ is the surface tension coefficient, $\delta$ is a Dirac delta function located at the interface, $\kappa$ is the curvature of the surface, and $\boldsymbol{n_i}$ is the normal vector of the liquid-gas interface and $\nabla_s$ is the surface gradient operator.

The interface normal is given by:

$$\boldsymbol{n_i} = \frac{\nabla \phi}{|\nabla \phi|} \quad (7)$$

The Dirac delta function is given by:

$$\delta = 6 |\nabla \phi| |\phi(1-\phi)| \quad (8)$$

The curvature of the surface is given by:

$$\kappa = -\nabla.\boldsymbol{n_i} \quad (9)$$

The surface gradient operator is given by:

$$\nabla_s = (I - \boldsymbol{n_i}\boldsymbol{n_i^T})\nabla \quad (10)$$

Also, the fluid's angle with the wall while wetting is assumed to be 90°. This is a standard assumption while solving sloshing flows[16].

The properties must be expressed regarding the level set variable for the fluid domain inside the tank. Density is given as:

$$\rho = \rho_1 + (\rho_2 - \rho_1)\phi \quad (11)$$

Viscosity is given as:

$$\mu = \mu_1 + (\mu_2 - \mu_1)\phi \tag{12}$$

where indices 1 and 2 denote fluid 1 and fluid 2, which in our case is the ferrofluid and air, respectively. Similarly, to calculate the Kelvin body force, the magnetic properties of the fluids must be expressed in terms of the level set variable.

$$\vec{F_m} = \vec{M}.\nabla\vec{H} \tag{13}$$

where the magnetization $\vec{M}$ is given by the constitutive relation:

$$\vec{M} = \chi_m \vec{H} \tag{14}$$

where, $\chi_m$ is the magnetic susceptibility and is obtained from the level set function for the whole domain as:

$$\chi_m = \chi_{m1} + (\chi_{m2} - \chi_{m1})\phi \tag{15}$$

An external coil with 2000 turns (N) picks up the electromotive force from sloshing. The electromotive force is given as:

$$\epsilon = -N\frac{d\Phi}{dt} \tag{16}$$

where, $\Phi$ is the magnetic flux obtained by integrating the magnetic flux density over the area.

d) Structural mechanics and piezoelectric:

$$\rho\frac{\partial^2 u_{solid}}{\partial t^2} = \nabla.\sigma + \vec{F_V} \tag{17}$$

where $u_{solid}$ is the displacement of the piezoelectric bimorph, $\sigma$ is the Cauchy stress tensor, and $\vec{F_V}$ is the external volumetric body forces acting upon the beam. The fluid-structure interaction stresses are caused by the reaction forces caused by the fluid on the bimorph. The force experienced by the fluid is calculated similarly and is then accounted for in mesh displacement to capture the effect of the fluid-structure interaction.

The strain on the bimorph causes electric displacement in the piezoelectric material and generates power. For our calculations, we have considered the piezo material 'z-x' or '3-1' as the material axis. The following equations describe the piezoelectric interaction:

$$S = s_E\, \sigma + d'\, E \tag{18}$$

$$D = d\, \sigma + \epsilon^T\, E \tag{19}$$

where, $S$ is the strain tensor, $s_E$ the elastic compliance matrix, $d$ is the piezoelectric strain coefficient matrix, $D$ is the electric displacement, $\epsilon$ the dielectric permittivity matrix, $E$ the electric field and $\sigma$ the stress tensor.

A commercially available ferrofluid developed by FerroTec Corporation, named **EFH – 3** is used in the study. The two piezoelectric materials used for the bimorph are PZT-7A and PVDF. The properties of the three materials are listed in Appendix C.

## 4. Numerical Modeling

The governing equations are solved in COMSOL Multiphysics using its inbuilt modules and coupling interfaces for fluid-structure interaction, separated two-phase flow with level set, piezoelectric effect, and internal wetted wall. **Figure 3** illustrates the meshing scheme used in COMSOL. The beam is meshed with very fine elements to track the movement of the mesh and Laplace mesh smoothing is used to move the mesh according to the velocity profile of the solid[20].

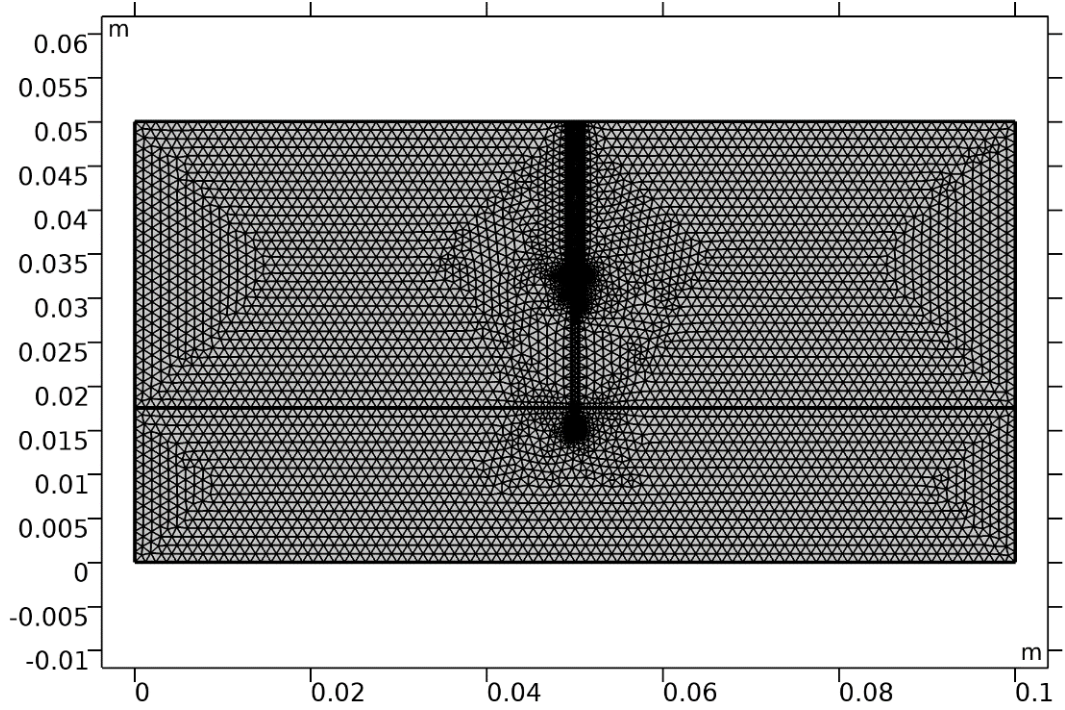

**Figure 4** Meshing of the domain of Energy harvester

In order to analyze the performance of the hybrid energy harvester, a combination of 5 different parameters are varied, as shown in **Table 1**. For all these geometrical parameters, the harvester is subjected to a frequency sweep at low frequencies, i.e. from 0.8 Hz to 1.2 Hz. The open circuit voltages are calculated using piezoelectric and sloshing equations.

**Table 1** Simulation cases for energy harvester

| Case | PVDF | PZT-7A |
|---|---|---|
| 1 | $X_0 = L, H = \frac{L}{2}, \frac{h}{H} = 0.35, \frac{h_b}{L} = 0.7, h_p = \frac{h_b}{4}$ | $X_0 = L, H = \frac{L}{2}, \frac{h}{H} = 0.35, \frac{h_b}{L} = 0.7, h_p = \frac{h_b}{4}$ |
| 2 | $X_0 = L, H = \frac{L}{2}, \frac{h}{H} = 0.35, \frac{h_b}{L} = 0.7, h_p = \frac{h_b}{2}$ | $X_0 = L, H = \frac{L}{2}, \frac{h}{H} = 0.35, \frac{h_b}{L} = 0.7, h_p = \frac{h_b}{2}$ |
| 3 | $X_0 = L, H = \frac{L}{2}, \frac{h}{H} = 0.4, \frac{h_b}{L} = 0.7, h_p = \frac{h_b}{4}$ | $X_0 = L, H = \frac{L}{2}, \frac{h}{H} = 0.4, \frac{h_b}{L} = 0.7, h_p = \frac{h_b}{4}$ |
| 4 | $X_0 = L, H = \frac{L}{2}, \frac{h}{H} = 0.4, \frac{h_b}{L} = 0.7, h_p = \frac{h_b}{2}$ | $X_0 = L, H = \frac{L}{2}, \frac{h}{H} = 0.4, \frac{h_b}{L} = 0.7, h_p = \frac{h_b}{2}$ |

The different physics interfaces are handled in COMSOL, using a single segregated iteration, with different steps solving for the dependent variables which are present in the step. The

steps have the following dependent variables and are executed in this sequential order for each time step:

a) Velocity and pressure

b) Level-set variable

c) Displacement field for the solid

d) Moving mesh

e) Piezoelectric voltage and charge

These equations are solved using the in-built Streamline Upwind Petrov Galerkin (SUPG) method, which takes care of the stability of the solution[21]. Since this is an unsteady problem, the algorithm needs to solve for time marching of these variables and not just a steady-state solution. Hence, for the solution to progress, all of the above steps need to be solved at each step, and once the residual goes below the set tolerance, the solution converges at that particular time. Furthermore, the time-stepping used is implicit adaptive time stepping, which adapts the time-step according to the non-linearity faced in each time step. The time marching method used here is the backward differencing formula (BDF) of order 2 or 1. Since the solver is implicit and adaptive, to obtain time step convergence, the maximum time step parameter is changed. A parameter of 0.005 sec was used as it was stable and much less expensive than a smaller time step.

Moreover, the PARDISO and MUMPS direct solvers are used to solve each segregated step to obtain convergence at each time step. Finally, grid convergence studies were also performed by altering the meshing parameter. In this case, the maximum mesh dimension was altered, and a significantly fine mesh was generated to capture the physics correctly.

## 5. Results & Discussion

**Figure 5** Flow Evolution for a) PZT7A and b) PVDF for f = 1 Hz at t = 0.47 sec represents a flow solution for the two piezos with an excitation frequency of 1 Hz at a time equal to 0.47 seconds. These results are for case 2, where the piezoelectric material covers half the substrate. The RMS voltages from the two configurations for all four cases are shown in **Figure 6**. The piezoelectric energy harvesting circuit and the ferrofluid sloshing energy harvesting circuit are separate from each other. This approach was used to understand the true potential of the proposed system and obtain design guidelines to transform it into practical designs. The observed voltages from the piezos are different from each other. The voltage from PZT 7A is approximately half of that from PVDF. This can be attributed to the order of magnitude lower stiffness of PVDF. Also, the simulations are run for 4 time periods, and the RMS voltage is calculated from the last three time periods (or cycles) data for the voltage.

Appendix B shows the voltage evolution plots for all the cases and for both the piezoelectric materials. Like the external excitation function, the voltage evolves as a periodic function.

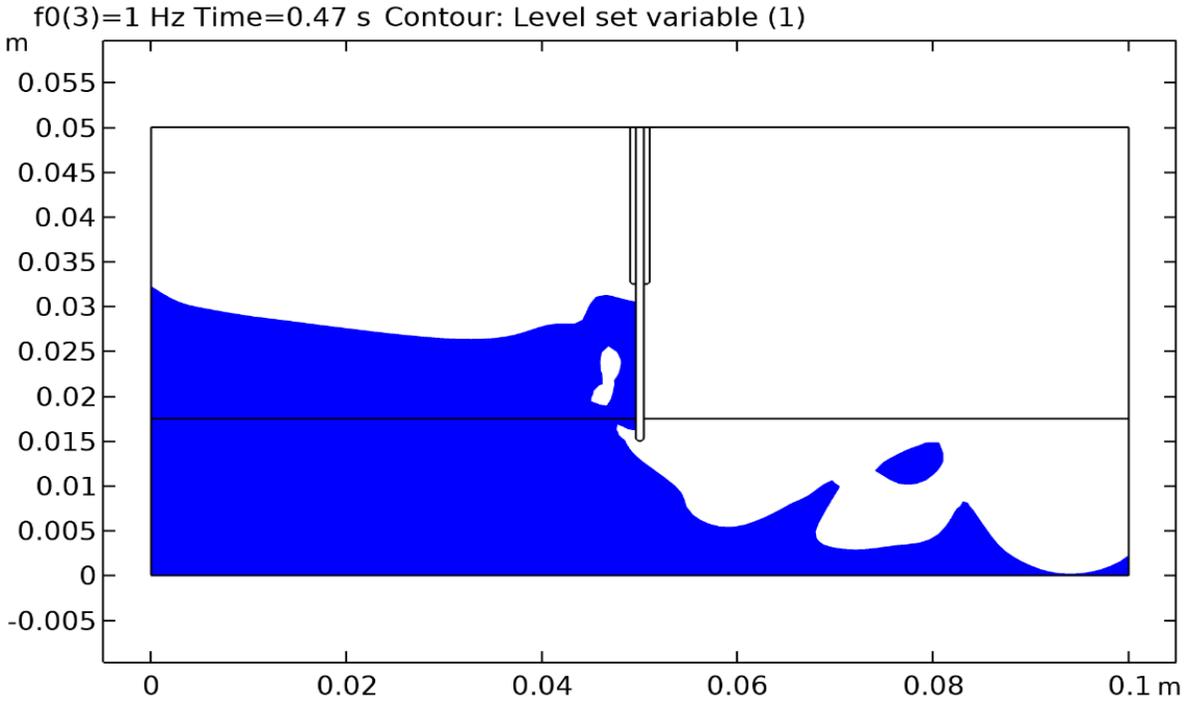

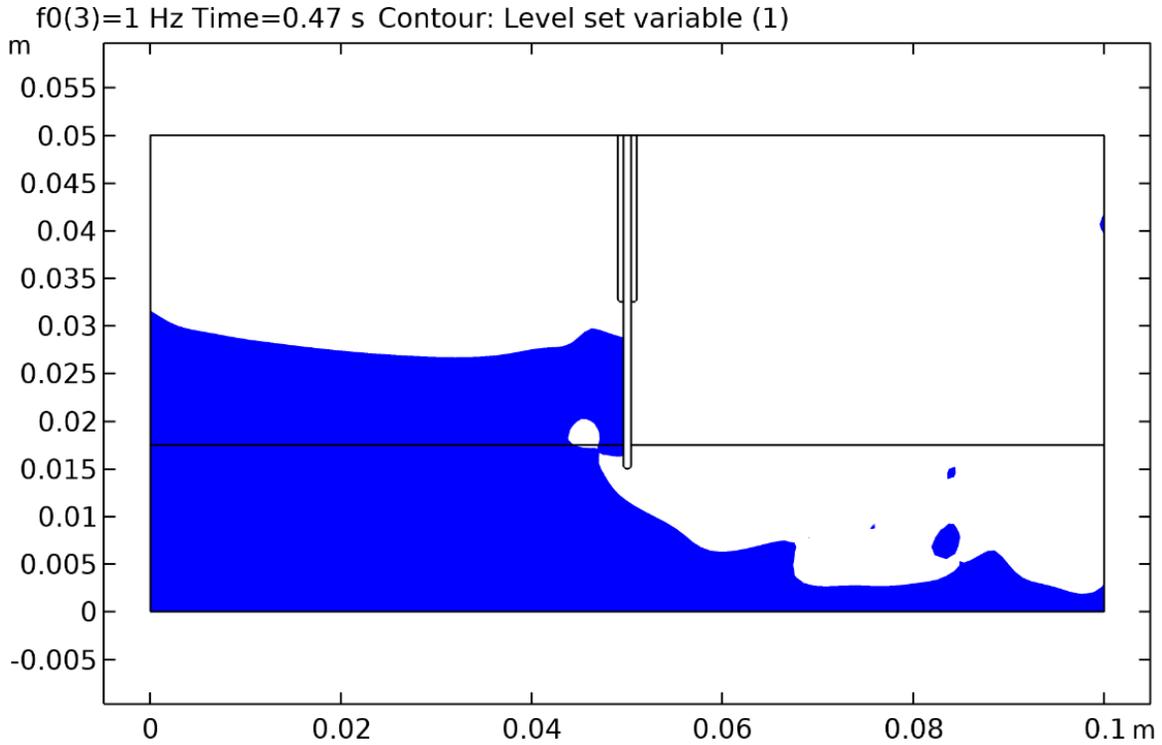

**Figure 5** Flow Evolution for a) PZT7A and b) PVDF for f = 1 Hz at t = 0.47 sec

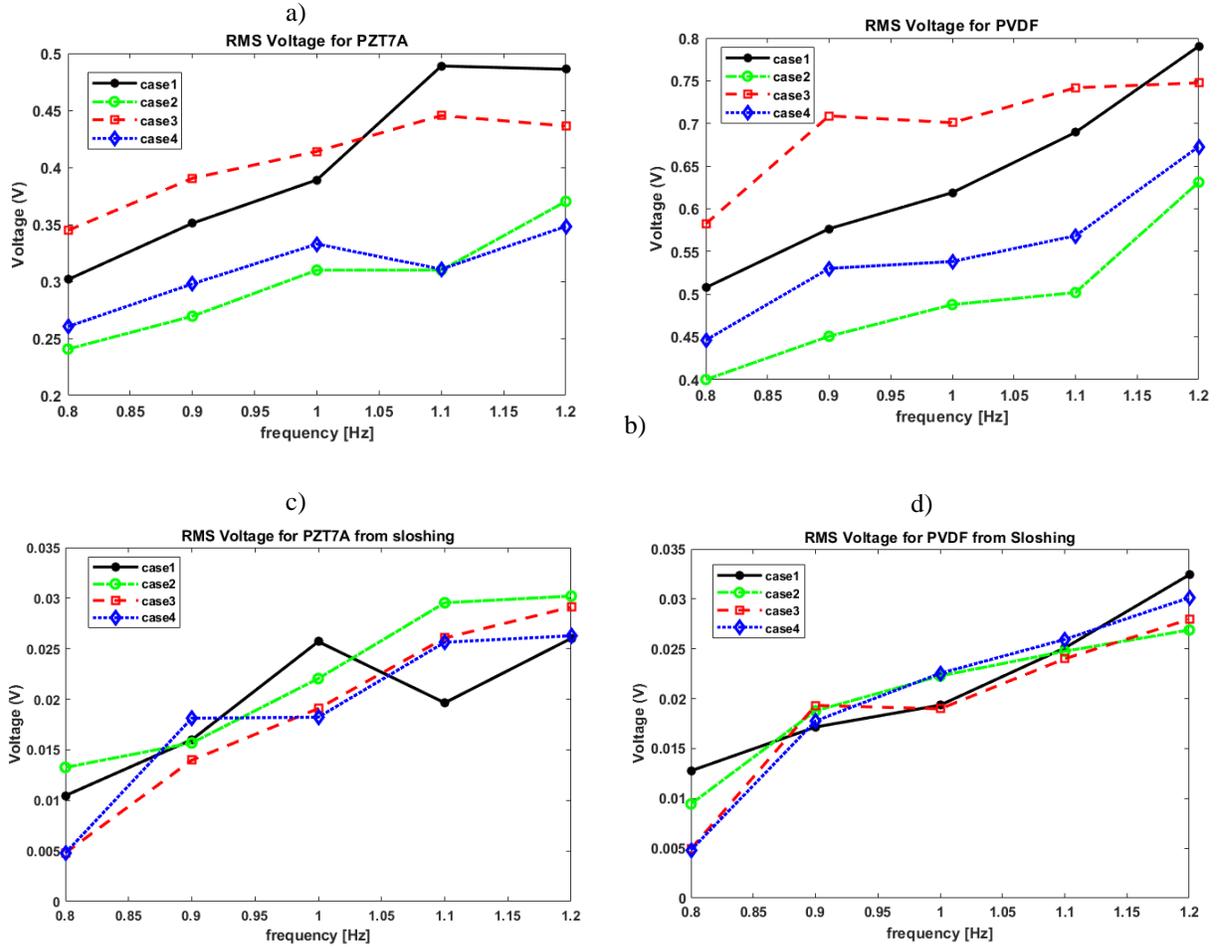

**Figure 6** RMS total voltage for a) PZT-7A and b) PVDF and Voltage from sloshing from c) PZT-7A and d) PVDF

Since the voltages from sloshing are very low and PVDF produces the highest voltage, two more magnetic field functions are introduced. Case 1 using the PVDF is repeated with these two new magnetic field functions for five values of maximum magnetic field, varying from 10 kA/m to 24 kA/m. The two magnetic field functions are shown in Error! Reference source not found.**6**.

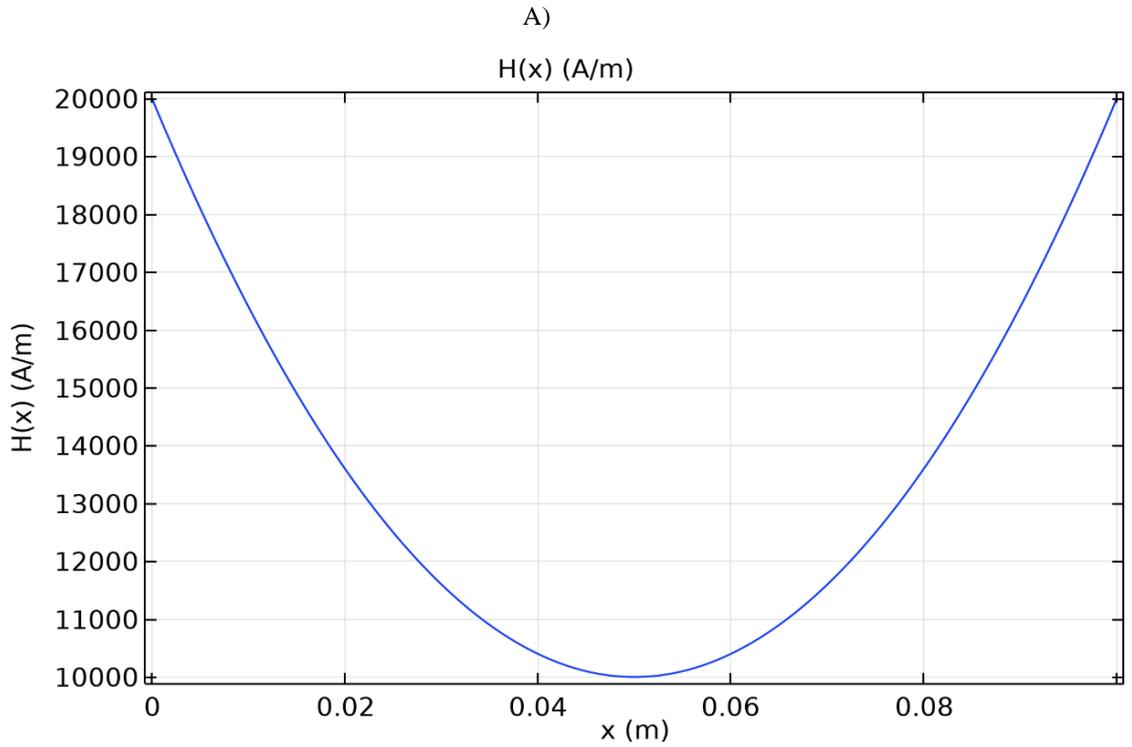

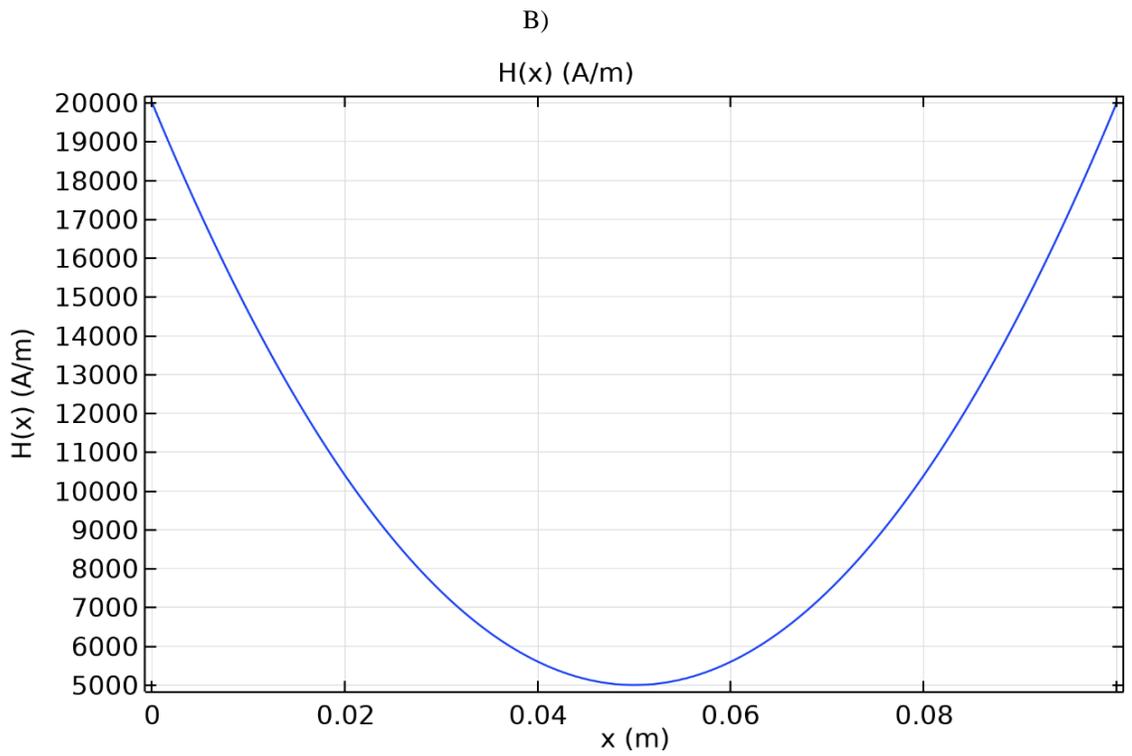

**Figure 7** Graph for a) Magnetic field function 1 and b) Magnetic field function 2, for H_max = 20[kA/m]

The RMS voltages from both the magnetic field functions for case 1 are shown in Error! Reference source not found.. The RMS voltage for PVDF is highest for the 10 kA/m case for both magnetic field functions. This can be attributed to an increase in the magnitude of the Kelvin body force, which by virtue of the two function shapes, will try to attract the ferrofluid towards the walls. Additionally, the voltage generated from sloshing increases with increasing maximum magnetic field.

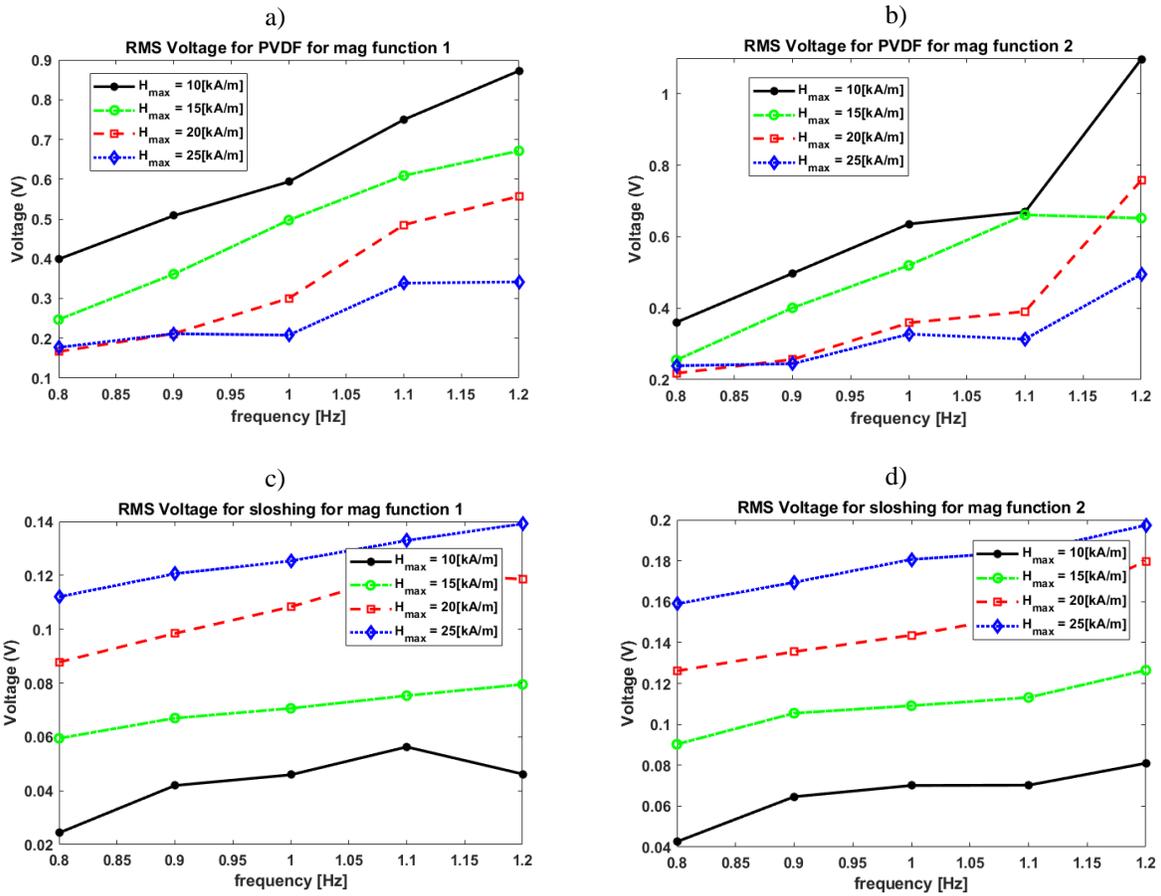

**Figure 8** Total RMS voltage for PVDF for a) Magnetic function 1 b) magnetic function 2 and sloshing voltage for PVDF for a) magnetic function 1 and b) magnetic function 2

Finally, in order to define the feasibility of the system, the output power must be optimized. In this chapter the output power from the piezo is prioritized in the sense that the parameters controlling the output power from piezo are embedded in the structure of the bimorph. Since, the simulation of the piezo was assumed to have very low coupling, the optimal impedance or resistance can be obtained by considering a simple AC circuit, where the optimal resistance value is given as[22], [23], [24]:

$$R_{opt} = \frac{1}{C_0 \omega}$$

where, $C_0$ is the parasitic capacitance of the piezoelectric and is given by:

$$C_0 = \frac{\epsilon_{33}^S A}{L}$$

where A is the area of the piezo and L the length of the piezo. The area is given as:

$$A = t_p W$$

where W is the width of the piezoelectric material. Calculating this from the material properties of PVDF and PZT-7A, the power comes out in nanowatts. This is because the natural frequency of vibration of the bimorph is very high as compared to the excitation frequency used here.

## 6. Conclusion

A feasibility study on the design of a hybrid piezoelectric-ferrofluid VEH is presented. To simplify the analysis, several factors were relaxed. Notably, the magnetic field was imposed as a simple function rather than a defined magnetic field by placing external magnets. Also, the bimorph was assumed to be thick enough not to disrupt the flow field, and hence, fluid loading on the structure becomes more significant. These factors helped establish the feasibility of a hybrid system such as this. However, an optimized system could not be established (i.e., the power output from the bimorph is very low). However, it can still be utilized for some applications which require low power. It must be noted that the present system can be established as a new alternative for renewable and seamless power. However, extensive optimization studies must be performed, and the optimized system capabilities must be established.

# Appendix A: Voltage Evolution plots for different cases

## 1. For PZT7A

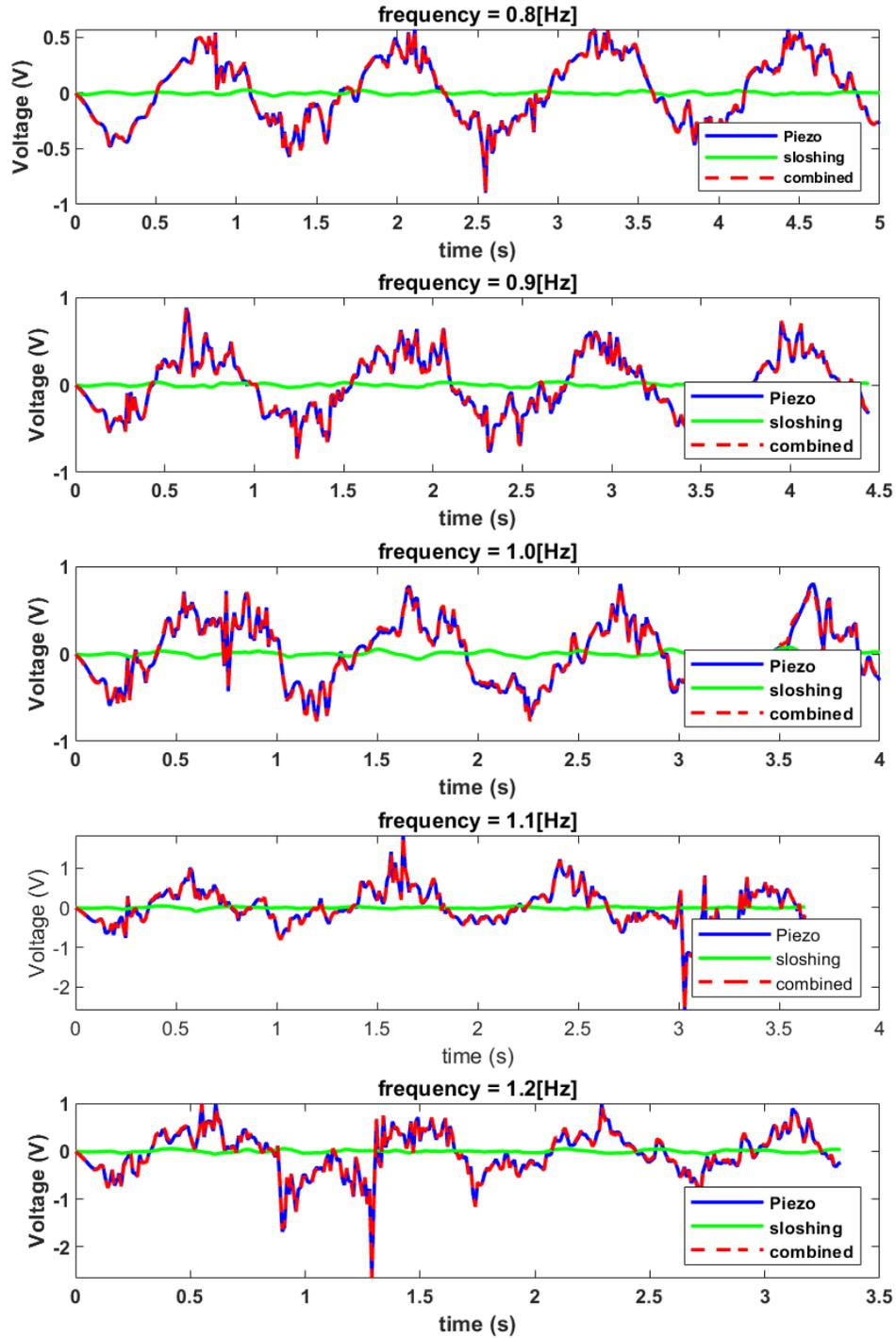

**Figure 9** Case1 time evolution for different frequencies with PZT7A

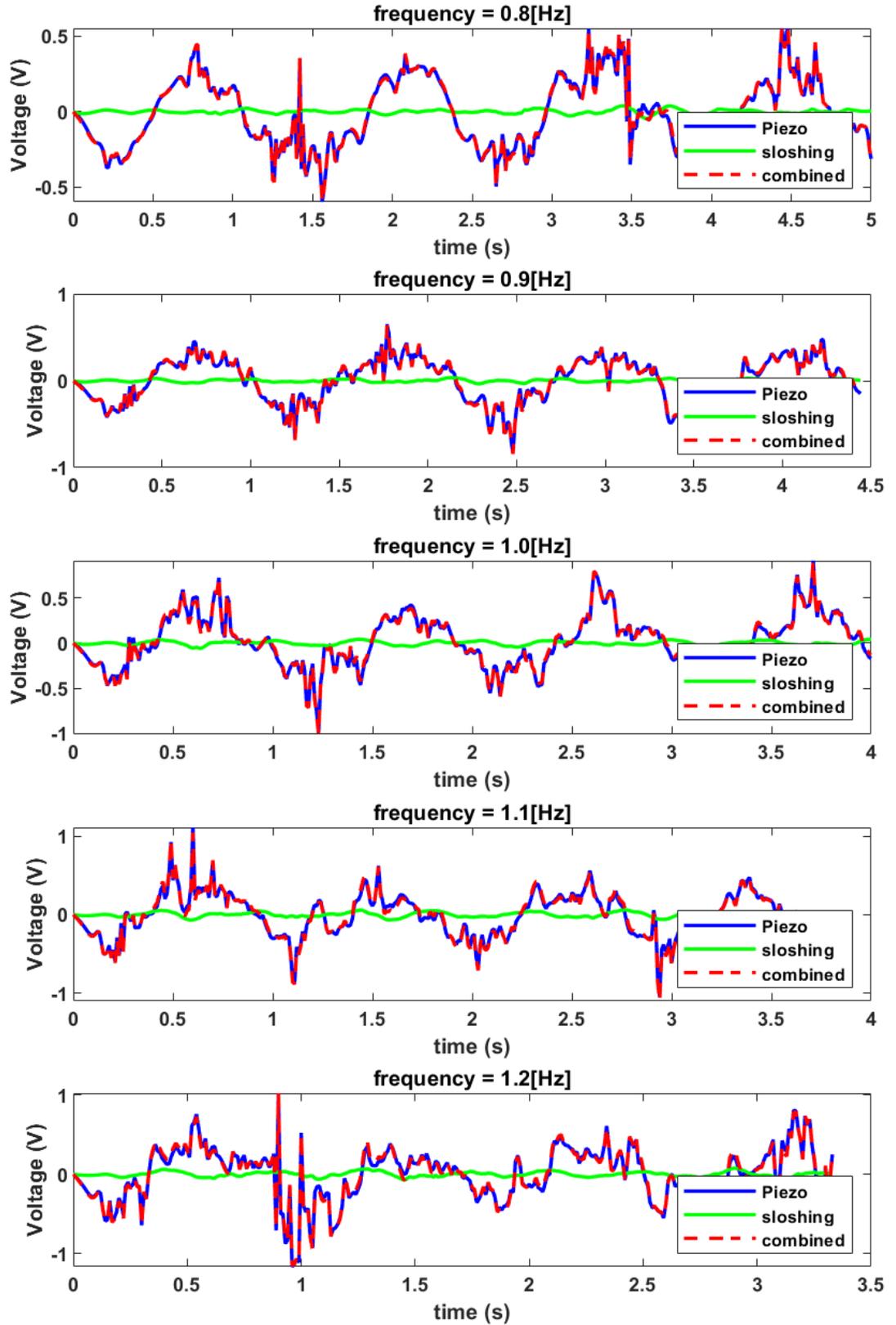

**Figure 10** Case2 time evolution for different frequencies with PZT7A

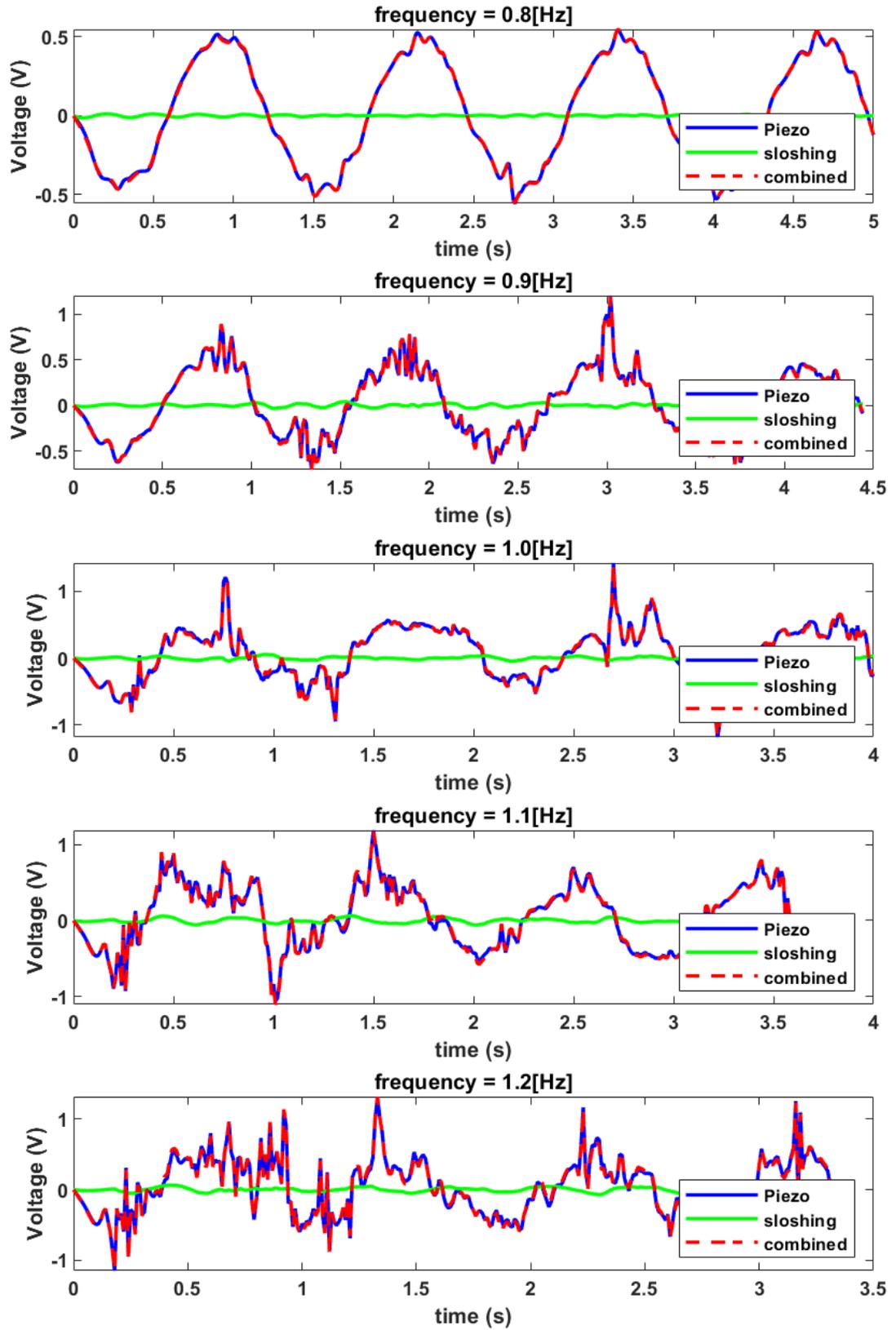

**Figure 11** Case3 time evolution for different frequencies with PZT7A

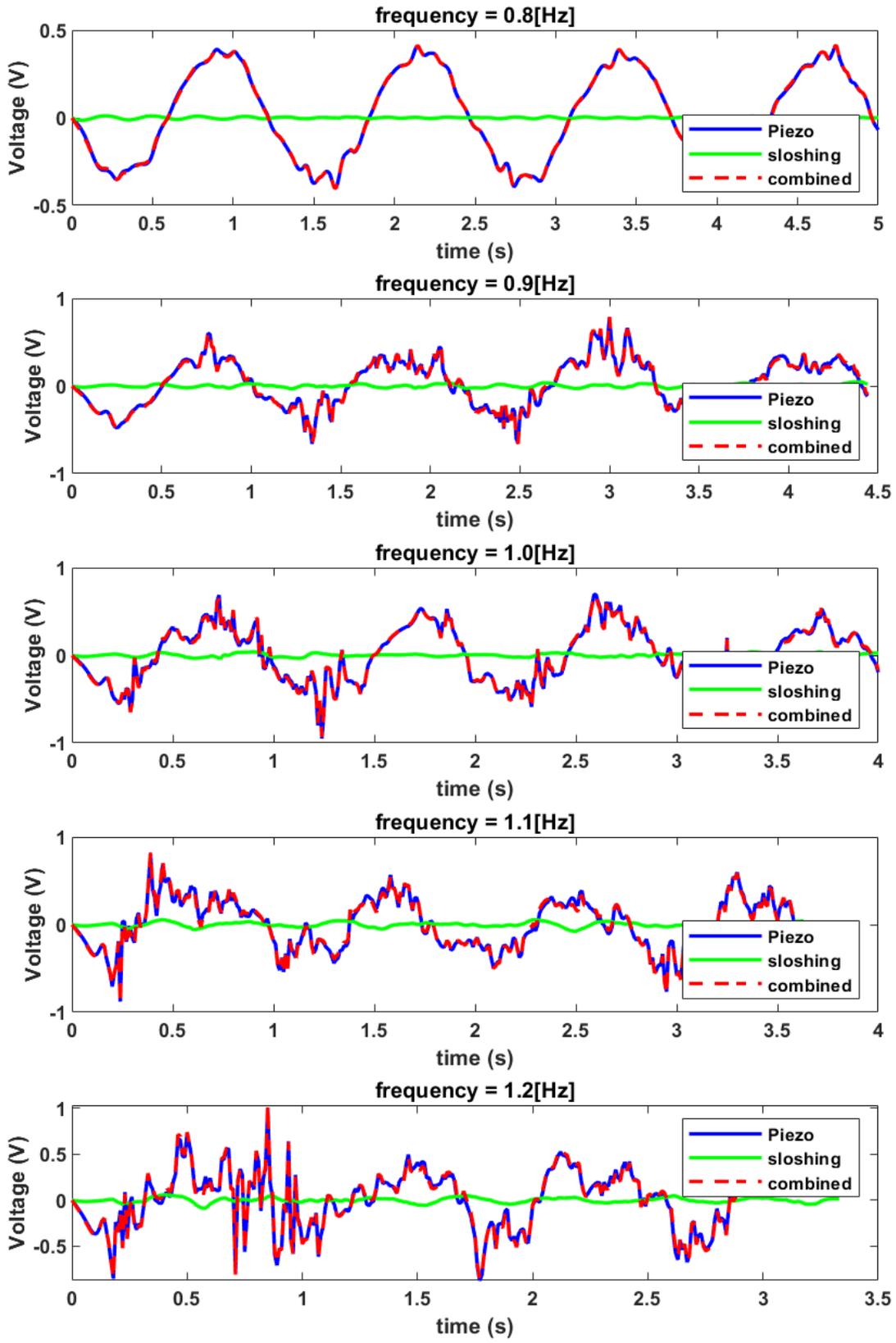

**Figure 12** Case4 time evolution for different frequencies with PZT7A

2. For PVDF

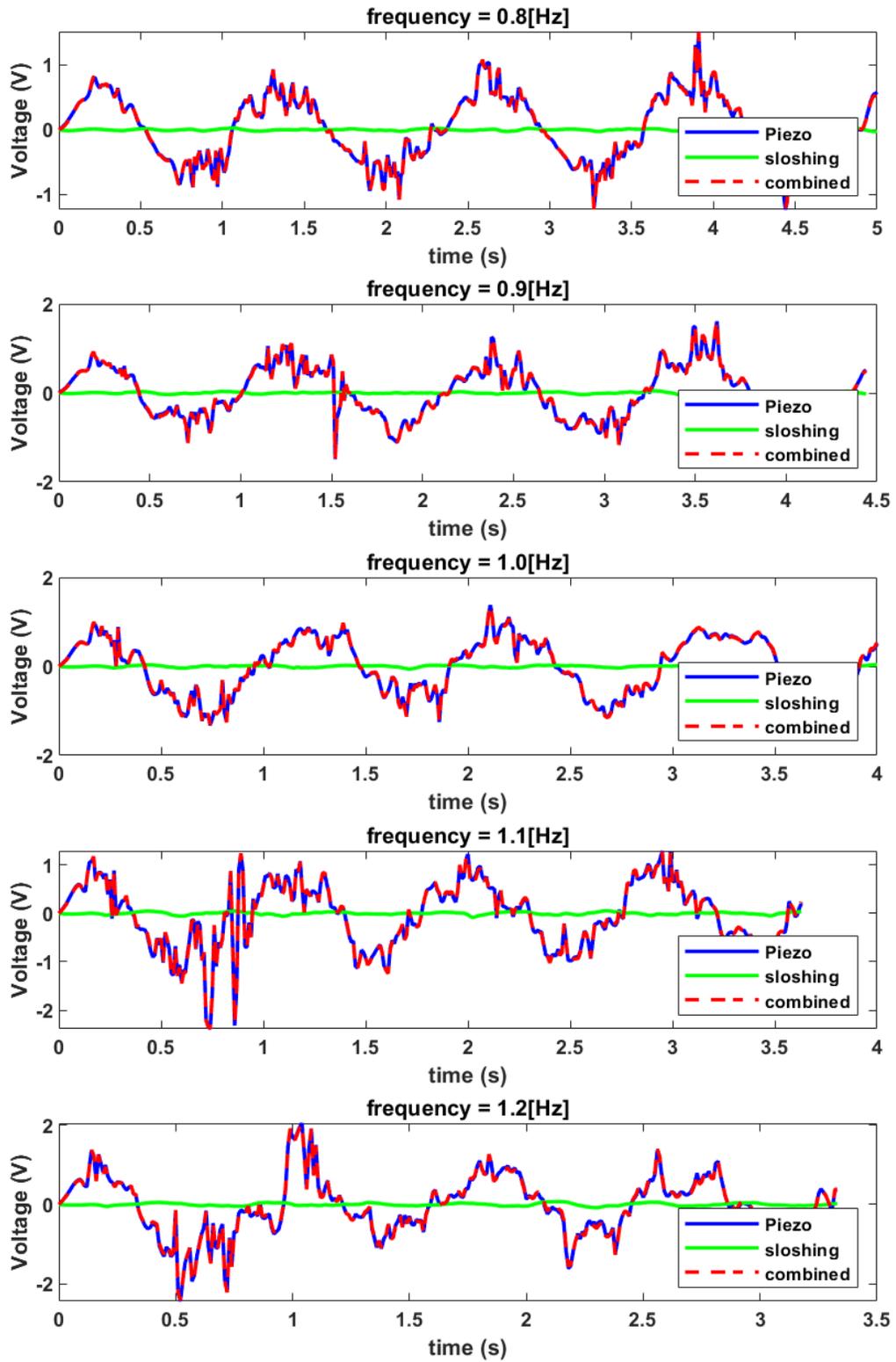

**Figure 13** Case1 time evolution for different frequencies with PVDF

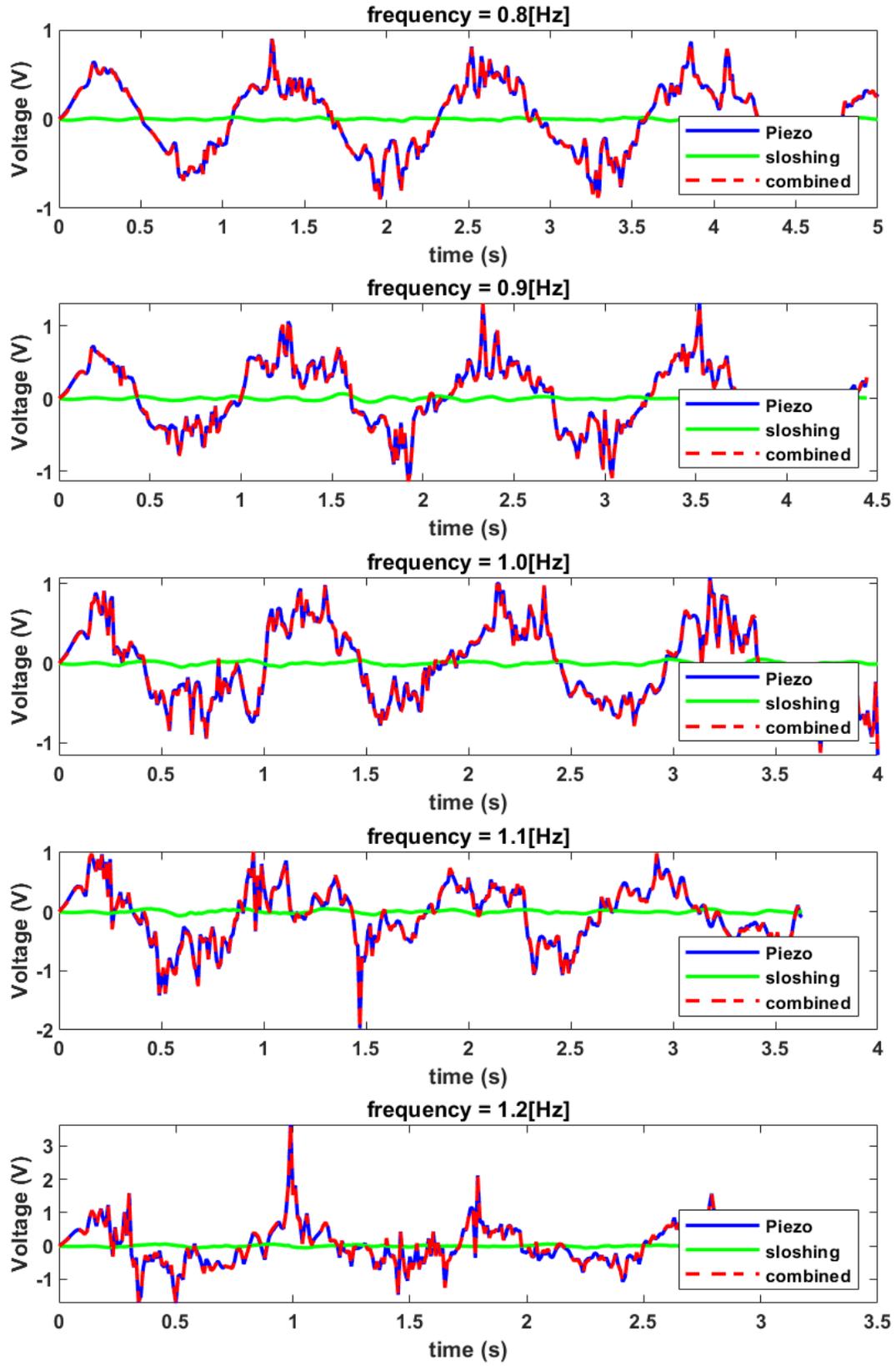

**Figure 14** Case2 time evolution for different frequencies with PVDF

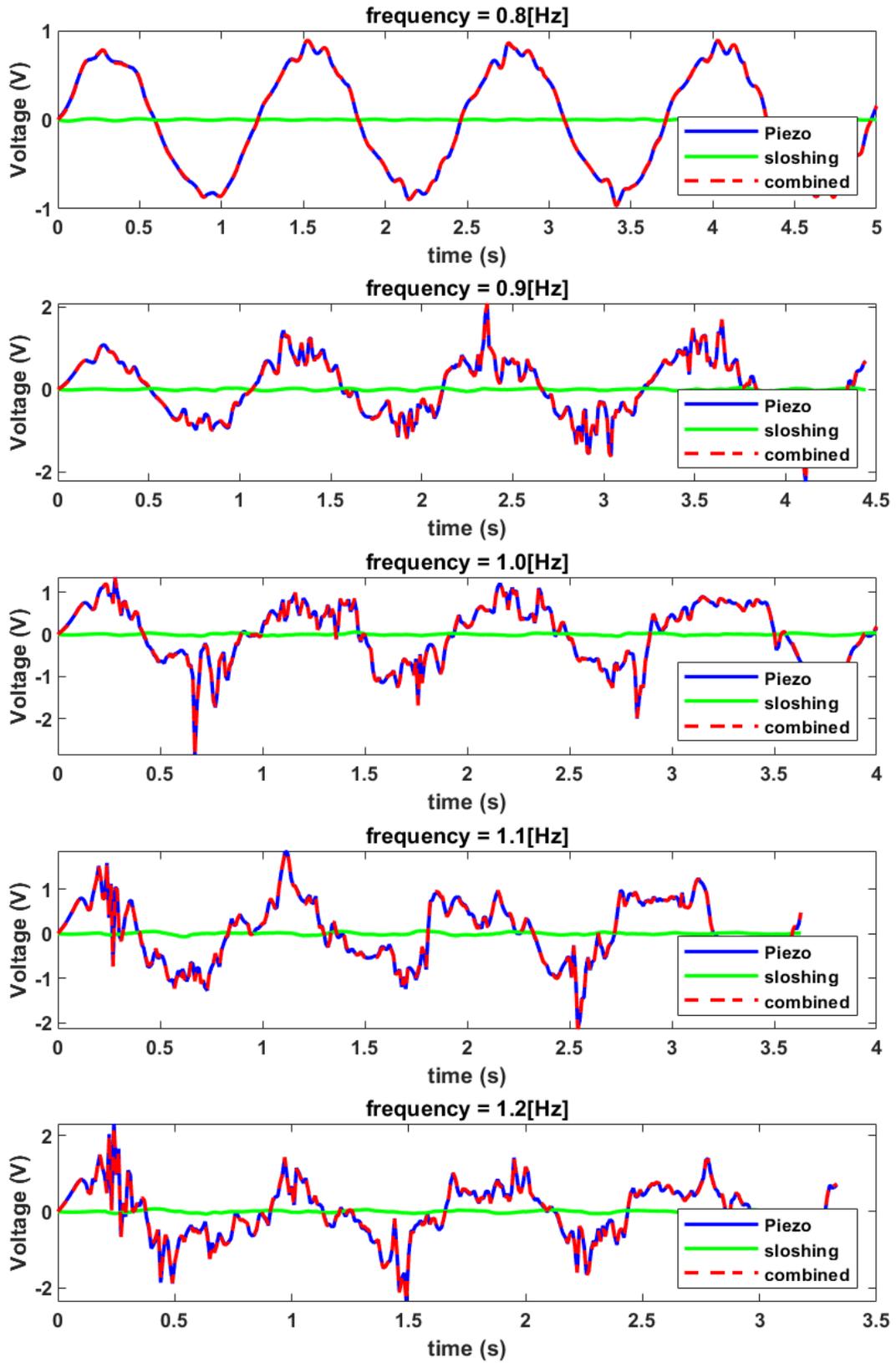

**Figure 15** Case3 time evolution for different frequencies with PVDF

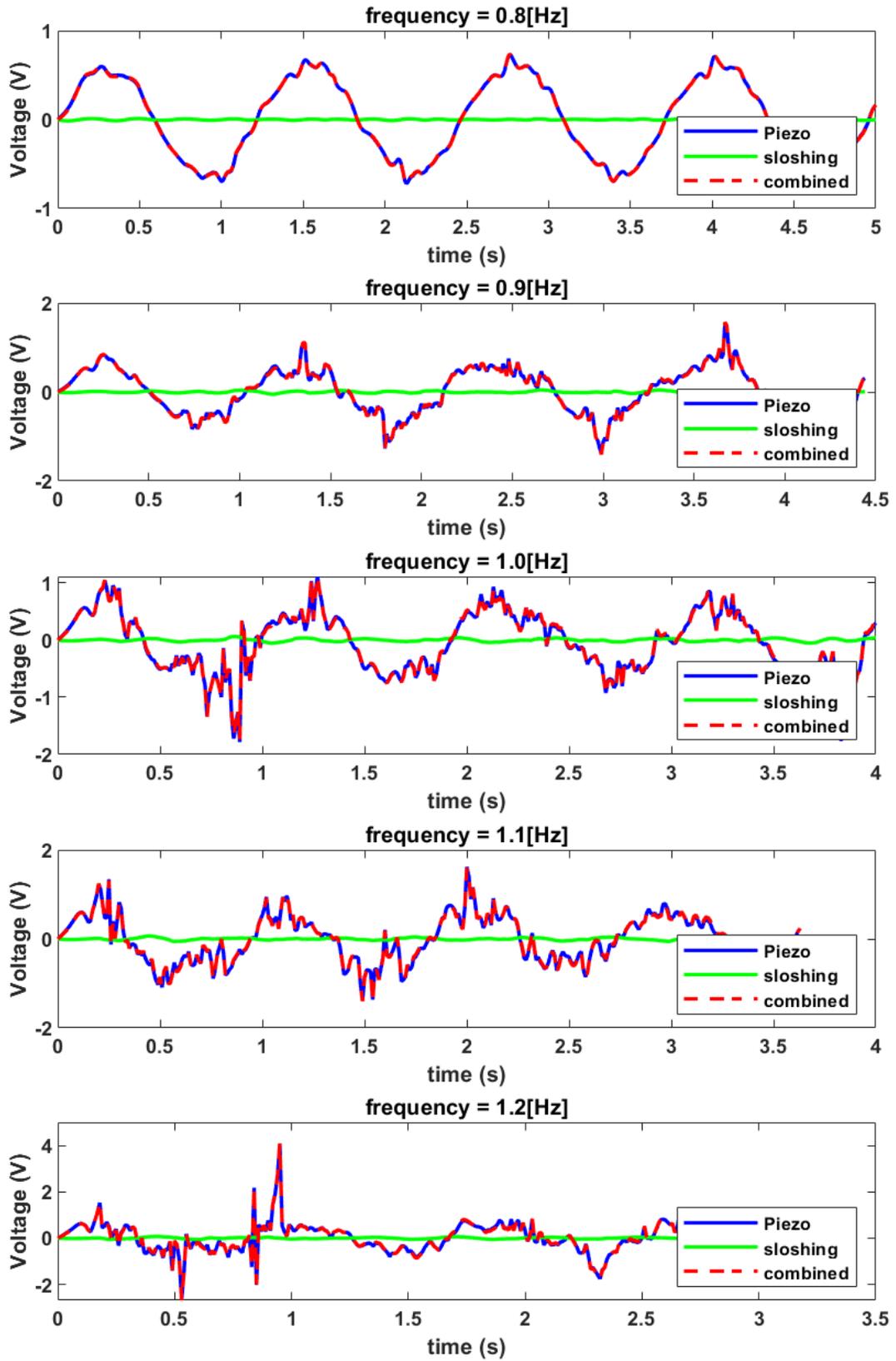

**Figure 16** Case4 time evolution for different frequencies with PVDF

## Appendix B: Physical properties of materials

Properties of EFH3 FerroTec corporation:

| Property | Value |
|---|---|
| Viscosity | 12 [mPa-s] |
| Density | 1420 [kg/m³] |
| Surface Tension | 24.15 [mN/m] |
| Magnetic susceptibility | 3.52 |
| Saturation magnetization | 65 [mT] |

Properties of Lead Zirconate Titanate (PZT-7A) are as follows:

The elastic compliance matrix is given by:

$$s_E = \begin{bmatrix} 10.7 & -3.58 & -4.6 & 0.00 & 0.00 & 0.00 \\ -3.58 & 10.7 & -4.6 & 0.00 & 0.00 & 0.00 \\ -4.6 & -4.6 & 13.9 & 0.00 & 0.00 & 0.00 \\ 0.00 & 0.00 & 0.00 & 34.0 & 0.00 & 0.00 \\ 0.00 & 0.00 & 0.00 & 0.00 & 34.0 & 0.00 \\ 0.00 & 0.00 & 0.00 & 0.00 & 0.00 & 28.6 \end{bmatrix} \times 10^{-12} \left(\frac{1}{Pa}\right)$$

The strain coefficient matrix is given by:

$$d = \begin{bmatrix} 0.00 & 0.00 & 0.00 & 0.00 & 36.0 & 0.00 \\ 0.00 & 0.00 & 0.00 & 36.0 & 0.00 & 0.00 \\ -6.00 & -6.00 & 15.3 & 0.00 & 0.00 & 0.00 \end{bmatrix} \times 10^{-11} \left(\frac{C}{N}\right)$$

The dielectric permittivity matrix is given by:

$$\epsilon^T = \begin{bmatrix} 930 & 0.00 & 0.00 \\ 0.00 & 930 & 0.00 \\ 0.00 & 0.00 & 425 \end{bmatrix} \times \epsilon_0$$

Where, $\epsilon_0$ is the permittivity of free space.

## Appendix C: Physical properties of materials

Properties of Polyvinylidene Flouride (PVDF) are as follows:

The elastic compliance matrix is given by:

$$s_E = \begin{bmatrix} 3.78 & -1.48 & -1.72 & 0.00 & 0.00 & 0.00 \\ -1.48 & 3.78 & -1.72 & 0.00 & 0.00 & 0.00 \\ -1.72 & -1.72 & 10.9 & 0.00 & 0.00 & 0.00 \\ 0.00 & 0.00 & 0.00 & 14.3 & 0.00 & 0.00 \\ 0.00 & 0.00 & 0.00 & 0.00 & 11.1 & 0.00 \\ 0.00 & 0.00 & 0.00 & 0.00 & 0.00 & 11.1 \end{bmatrix} \times 10^{-10} \left(\frac{1}{Pa}\right)$$

The strain coefficient matrix is given by:

$$d = \begin{bmatrix} 0.00 & 0.00 & 0.00 & 0.00 & 0.00 & 0.00 \\ 0.00 & 0.00 & 0.00 & 0.00 & 0.00 & 0.00 \\ 13.58 & 1.476 & -33.8 & 0.00 & 0.00 & 0.00 \end{bmatrix} \times 10^{-12} \left(\frac{C}{N}\right)$$

The dielectric permittivity matrix is given by:

$$\epsilon^T = \begin{bmatrix} 7.4 & 0.00 & 0.00 \\ 0.00 & 9.3 & 0.00 \\ 0.00 & 0.00 & 7.74 \end{bmatrix} \times \epsilon_0$$

Where, $\epsilon_0$ is the permittivity of free space.